\documentclass[journal]{IEEEtran}
\ifCLASSINFOpdf
   \usepackage[pdftex]{graphicx}
   \graphicspath{{../pdf/}{../jpeg/}}
   \DeclareGraphicsExtensions{.pdf,.jpeg,.png}
\else
   \usepackage[dvips]{graphicx}
   \graphicspath{{../eps/}}
   \DeclareGraphicsExtensions{.eps}
\fi
\usepackage{amsmath}
\ifCLASSOPTIONcompsoc
  \usepackage[caption=false,font=normalsize,labelfont=sf,textfont=sf]{subfig}
\else
  \usepackage[caption=false,font=footnotesize]{subfig}
\fi
\usepackage{fixltx2e}
\usepackage{url}

\begin{document}
%
\title{An ultra-sensitive balanced detector with low noise for continuous-variable
quantum key distribution}
%
%
%

\author{Qi-Ming~Lu,
        Qi~Shen,
        Yuan Cao,
        Sheng-Kai Liao
        and~Cheng-Zhi~Peng
\thanks{This work was supported by the National Natural Science Foundation of China under Grant 11705191, the Anhui Provincial Natural Science Foundation under Grant 1808085QF180, and the Natural Science Foundation of Shanghai under Grant 18ZR1443600.}
\thanks{Authors are with Hefei National Laboratory for Physical Sciences at the Microscale and Department of Modern Physics, University of Science and Technology of China, Hefei 230026, China; and Chinese Academy of Sciences (CAS) Center for Excellence and Synergetic Innovation Center in Quantum Information and Quantum Physics, University of Science and Technology of China, Shanghai 201315, China(e-mail:luqiming@mail.ustc.edu.cn); (e-mail:shenqi@ustc.edu.cn); (e-mail: skliao@ustc.edu.cn).}}

\maketitle

\begin{abstract}
A well-balanced detector with high sensitivity and low noise is presented in this paper. The two-stage amplification structure is used to increase electronic gain while keeping an effective bandwidth of about 70 MHz. In order to further reduce electronic noise, a junction field-effect transistor(JFET) is connected between photodiodes and transimpedance amplifier to reduce the impact of amplifier leakage current. Benefit from these designs, the root-mean-square(RMS) of noise voltage is about 6 mV with a gain of 3.2E5 V/W, and it means an ultra-low noise equivalent power density of 2.2E-12 W/rtHz, only half of common low-noise commercial detectors. In addition, two photodiodes in similar frequency response are selected for detector and make the common mode rejection ratio(CMRR) of detector reached 53 dB, about 13 dB higher than commercial detectors. Further tests indicate that 16.8 dB shot-noise to electronic-noise ratio is measured in our detector, which is  better than most high speed balanced detectors.
\end{abstract}


%
\IEEEpeerreviewmaketitle

\section{Introduction}
%
%
%
%
Quantum key distribution is of great significance for future information security, while
continuous-variable quantum key distribution (CVQKD) is an important implementation path.
Compared to a QKD system using single-photon detector, the CVQKD which based on
homodyne detection encodes random number into the amplitude and phase of the pulsed laser
in quantum level at the transmitter and extracts random number through a balanced detector at
the receiver[1]. The detection method of CVQKD is based on the interference between weak signal
light and a strong light with the same frequency and synchronized phase, usually called local
oscillator light. This method filters out the background light mixed in the signal light and ensures that it not only achieves higher key rates over short distances but also possesses the potential to communicate in daylight[2].
\par According to [5], requirement of CMRR is greater than 45 dB for a typical CVQKD experimental system, otherwise extra excess noise caused by the imbalance of detector will decrease key distribution rate. However, although the existing commercial balanced detectors are well established, they are not suitable for CVQKD because their CMRR is usually less than 40 dB. Simultaneously, the shot-noise to electronic-noise ratio of detector is another important factor to improve the rate, which usually need to be greater than 10 dB in order to achieve a better result[5], and commercial detectors usually don't pay attention to this performance. Therefore we developed a dedicated balanced detector for our CVQKD experiments, it has very low noise in a gain of 3.2E5 V/W to extract information from weak quantum-level signal light  in a high signal-to-noise ratio. This paper introduces our design and tests for the low noise sensitive balanced detector. The design and structure of  detector are described in Section \uppercase\expandafter{\romannumeral2} and the test results are displayed in Section \uppercase\expandafter{\romannumeral3}. After that Section \uppercase\expandafter{\romannumeral4} is a conclusion.
\section{Design and structure}
The balanced detector used in early CVQKD experiments converts photons to electrons
through photodiode and transforms it to voltage by charge-sensitive amplifier[3]. However, this method causes a long signal tail, which is no longer appropriate as the pulse repetition frequency in experiments increased from hundreds of
KHz to dozens of MHz. To solve the problem, we use transimpedance amplifier instead.
 \par When a signal light interferes with local oscillator light, the output has following relationship[6]
\begin{equation}\label{Pout}
\begin{split}
  &P_{out}\propto | A_Se^{-i(\omega_St+\phi_S)}+A_Le^{-i(\omega_Lt+\phi_L)}|^2\\
  & \approx P_S+P_L+2\sqrt{P_SP_L}cos[(\omega_S-\omega_L)t+(\phi_S-\phi_L)].
\end{split}
\end{equation}
$A$, $\omega$, $\phi$, $P$ are  the carrier amplitude, phase, frequency, power of signal and local oscillator light respectively. Since the balanced detector has photocurrent decreasing structure, so that the DC component $P_S$, $P_L$ is eliminated and the last item is retained. In CVQKD experiments, we usually use the same frequency signal and local oscillator, and lock the phase of local oscillator on the signal light, this means that the peak output of balanced detector is
\begin{equation}\label{Vout}
  V_{out}=2GR(\lambda)\sqrt{P_SP_L}.
\end{equation}
Where $G$ is the electronic gain of balanced detector and $R(\lambda)$ is the responsivity of photodiodes. For our photodiodes, the $R(\lambda)$ is typically 1 A/W at 5 V bias voltage. Commercial detectors typically use noise equivalent power density(NEP) to represent noise levels, the value is
\begin{equation}\label{NEP}
  NEP=\frac{V_{rms}}{GR(\lambda)\sqrt{f_{-3dB}}}.
\end{equation}
based on theory in [6]. From (2)(3) we can see that the higher the gain of detector is, the lower the NEP is while maintaining bandwidth and low noise, which also means a higher sensitivity[6]. Therefore a two-stage amplification circuit structure is used in detector to make it possible achieving an ultra-high sensitivity, so that weak quantum signals can be detected.
\par Two operational amplifiers OPA847 from TI are used in this circuit with an ultra-low equivalent noise voltage density of $0.85\ nV/\sqrt{Hz}$. In order to ensure sufficient sensitivity, the gain of transimpedance amplifier is set to 8K and the voltage amplification factor is 40 times, resulting in a total gain of 320K. Two specially selected InGaAs PIN photodiodes(FGA01FC, Thorlabs, worked in 1550 nm) are serially connected for photocurrent reduction and their 2 pF junction capacitances are helpful to increase detector bandwidth. A low noise JFET(BF862) is connected between photodiodes and transimpedance amplifier to interdict the loop of amplifier leakage current flowing through feedback resistor, so that the noise  is suppressed[4]. The entire balanced detector is made into a 60 mm x 75 mm four-layer printed circuit board and powered by ¡À6 V with 40 mA normal operating current of each power rail. Detailed structure is shown in Figure 1.

\begin{figure}[!t]
\centering
\includegraphics[width=3.2in]{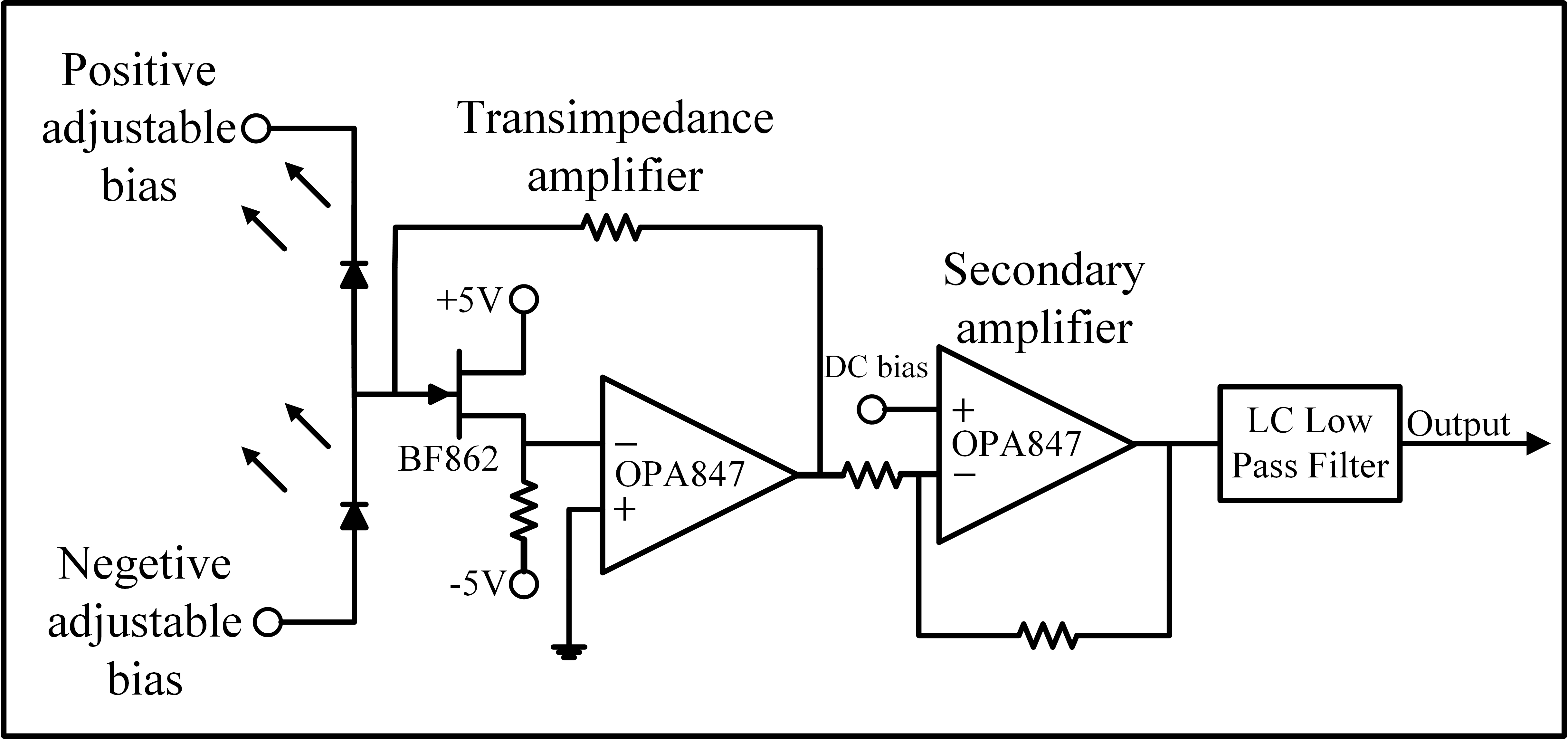}
\caption{Structure of detector.}
\label{Structure of detector}
\end{figure}
\par Due to the difference in frequency response of different PDs, for the same pulse laser, the shape of electrical signals output by these PDs will not be exactly the same. Thus a 5 ns pulse laser is used to compare the difference in response between different PDs, and the closest two of them are selected as inputs for the balanced detectors in order to achieve a higher CMRR level. Results of the comparison are shown in Figure 2, normalized Euclidean distances are used for this comparison. As the main reason for the difference in frequency response is the difference in junction capacitance between PDs[6], which can be slightly compensated by the offset voltage applied at both ends, so the bias voltage of photodiodes in detector are designed to be finely tuned to further improve the CMRR.

\begin{figure}[!t]
\centering
\includegraphics[width=3.2in]{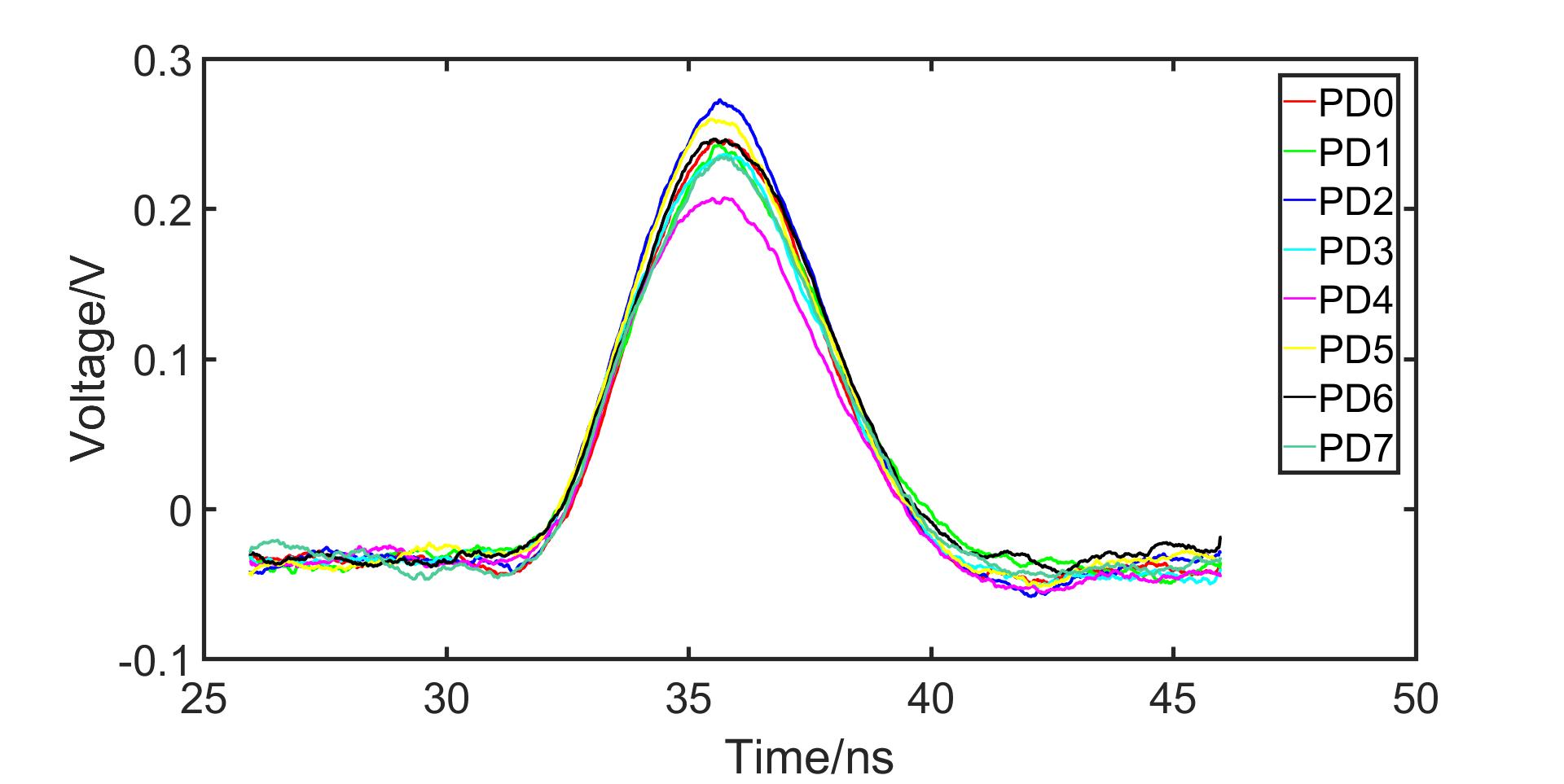}
\caption{Frequency response test of 8 PDs. The test is based on 5ns pulse laser.}
\label{Electronic noise of detector}
\end{figure}
\section{Test results}
We use the system shown in Figure 3 to test our balanced detector. Since the tests of CMRR and shot-noise to electronic-noise ratio don't need to interfere, so only the local oscillator beam injects to BS and the signal light is set to none(vacuum). Although the manual shows that responsivity of two photodiodes should be 1 A/W at 5 V bias, but in fact there are still slight differences between them. Simultaneously, the optical path is slightly different between two fibers from BS to balanced detector, so we must use variable optical attenuator (VOA) and optical delay line (ODL) to precisely control the delay and attenuation of two optical paths for achieving a high balanced test system.
\begin{figure}[!t]
\centering
\includegraphics[width=3.2in]{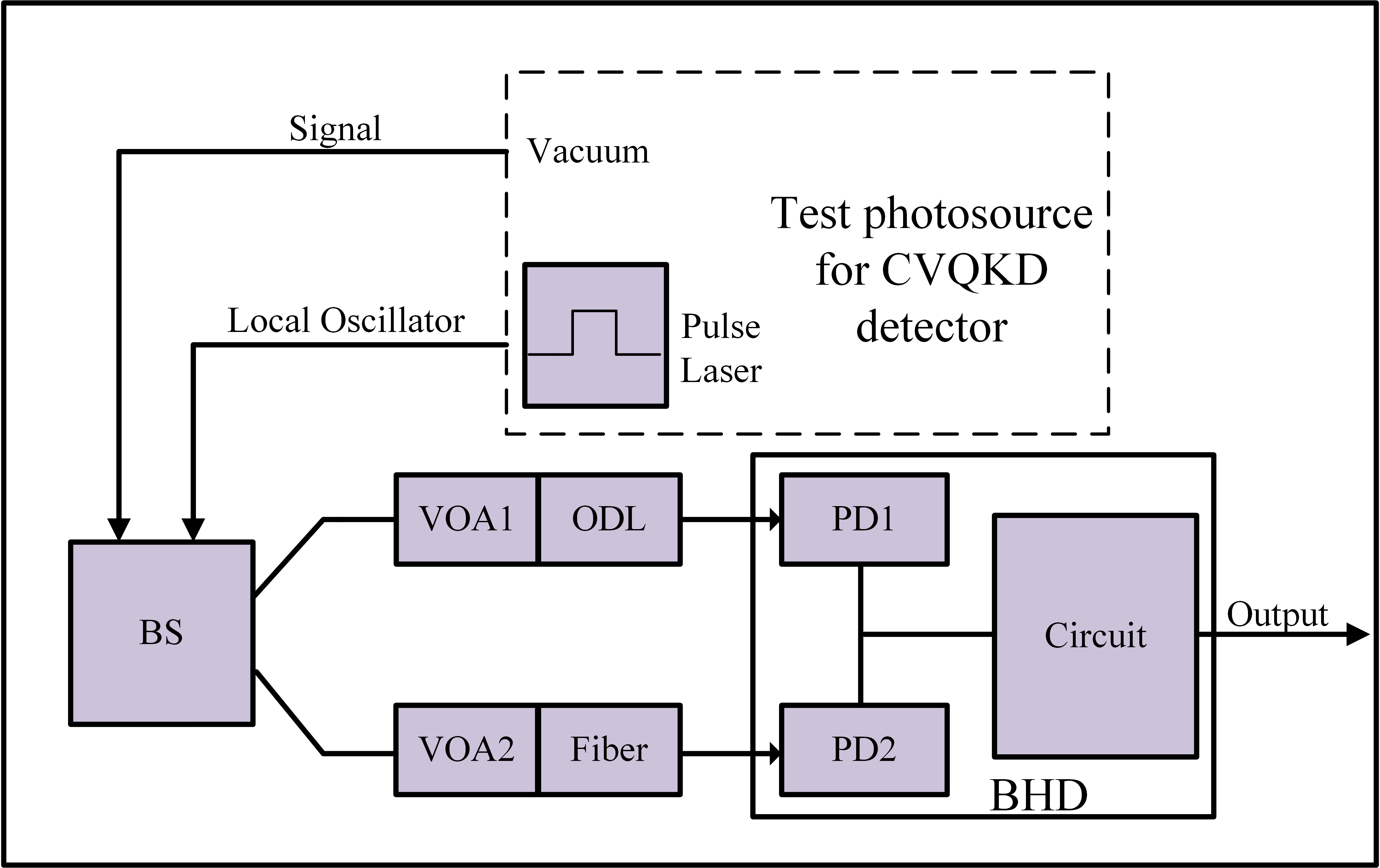}
\caption{Detector test system. Vacuum means no signal light input, the entire test is based on local oscillator light only.}
\label{Electronic noise of detector}
\end{figure}
\par Benefit from the JFET between photodiodes and transimpedance amplifier, the RMS of noise voltage is about 6 mV, and the two-stage amplification circuit structure makes the gain reached 3.2E5 V/W while keeping an effective bandwidth of 70 MHz. Figure 4 shows the electronic noise and bandwidth test of detector, substitute the result into (3) and the NEP is about $2.2\ pW/\sqrt{Hz}$.
\begin{figure}[!t]
\centering
\subfloat[]{\includegraphics[width=2.5in]{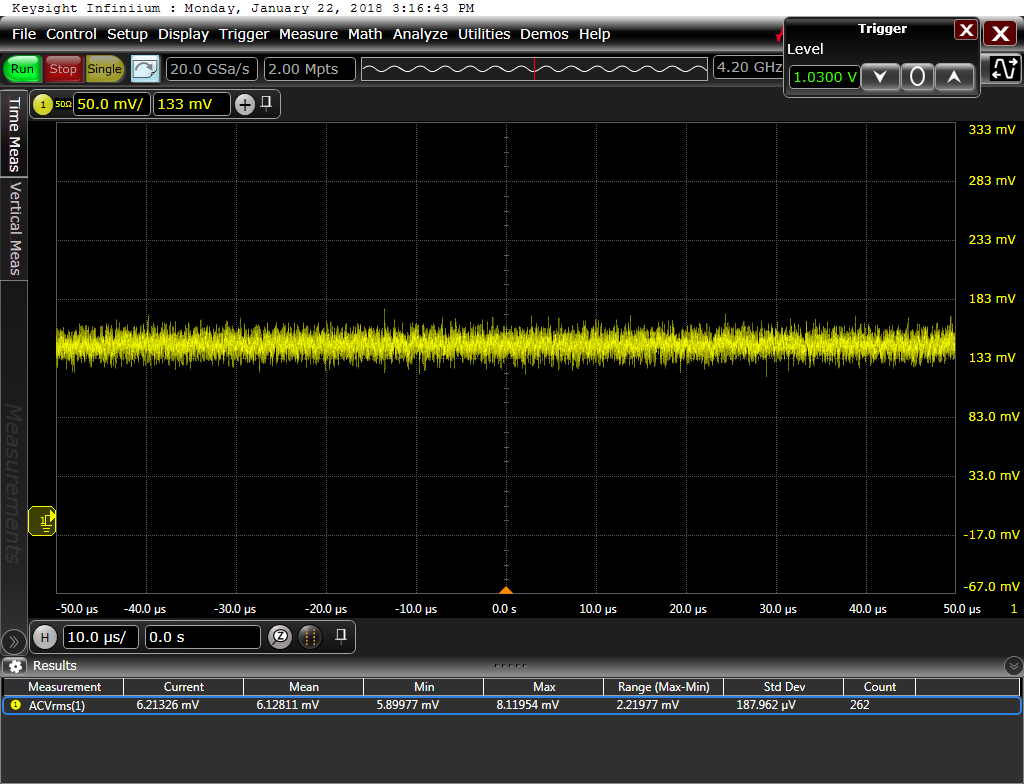}
\label{fig_first_case}}
\hfil
\subfloat[]{\includegraphics[width=3.2in]{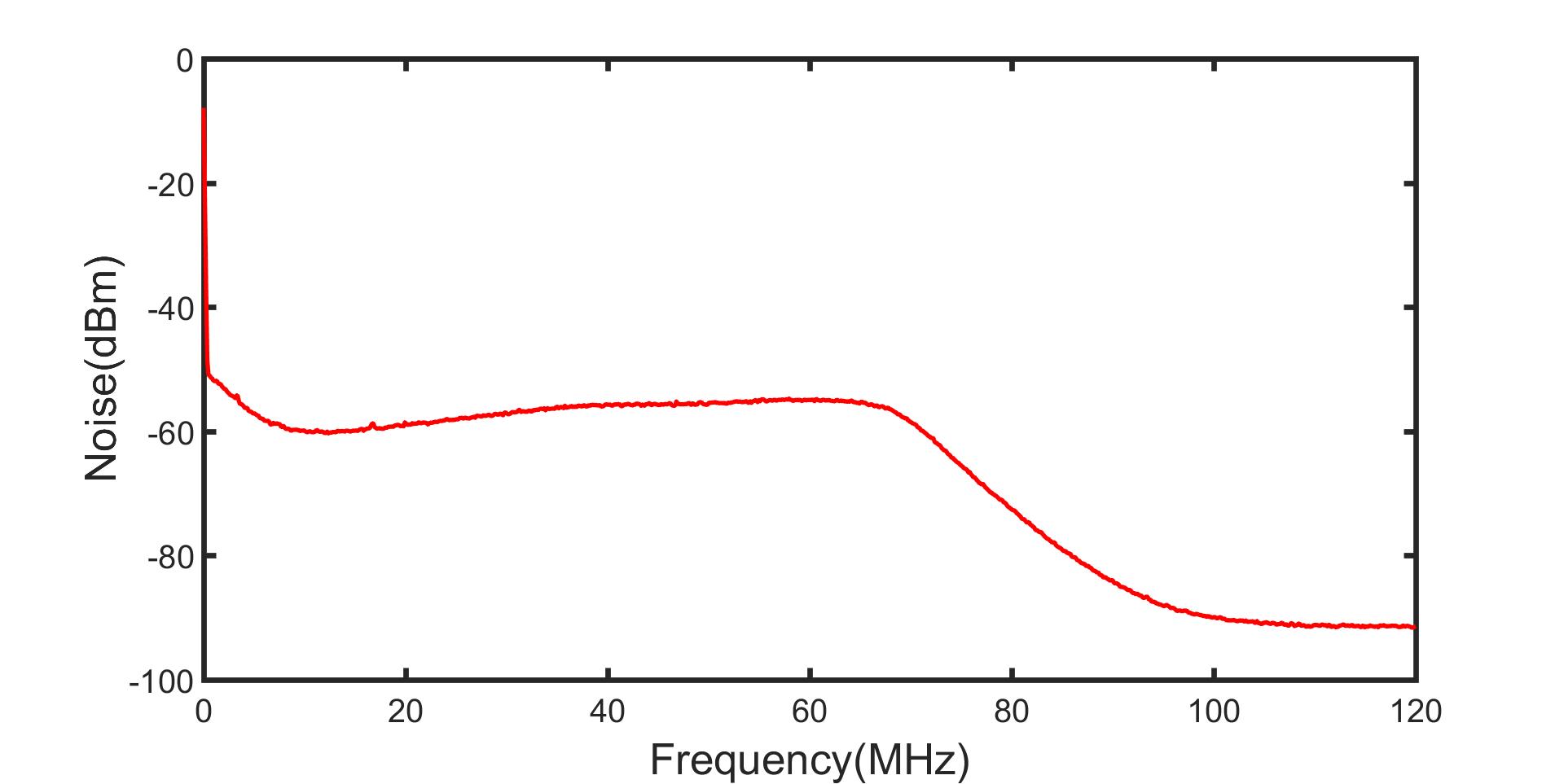}
\label{fig_second_case}}
\caption{Electronic noise and bandwidth test of detector. (a) is oscilloscope screenshot of detector electronic noise floor. The average RMS value is 6.128 mV with an oscilloscope noise floor of 0.6 mV, this means the detector noise is about 6 mV. (b) is the spectrum analyzer test of detector electronic noise floor, a 70 MHz -3dB bandwidth can be seen from the figure.}
\label{fig_sim}
\end{figure}
\par Figure 5 is the CMRR test results, which based on 40 MHz repetition rate pulse laser with 7.5 uw average power, and the results shows that CMRR eventually reaches 53 dB, about 13 dB higher than commercial detectors. This will be helpful to suppress extra noise due to detector imbalance and increase the key rate of QKD system[5].
\begin{figure}[!t]
\centering
\includegraphics[width=3.2in]{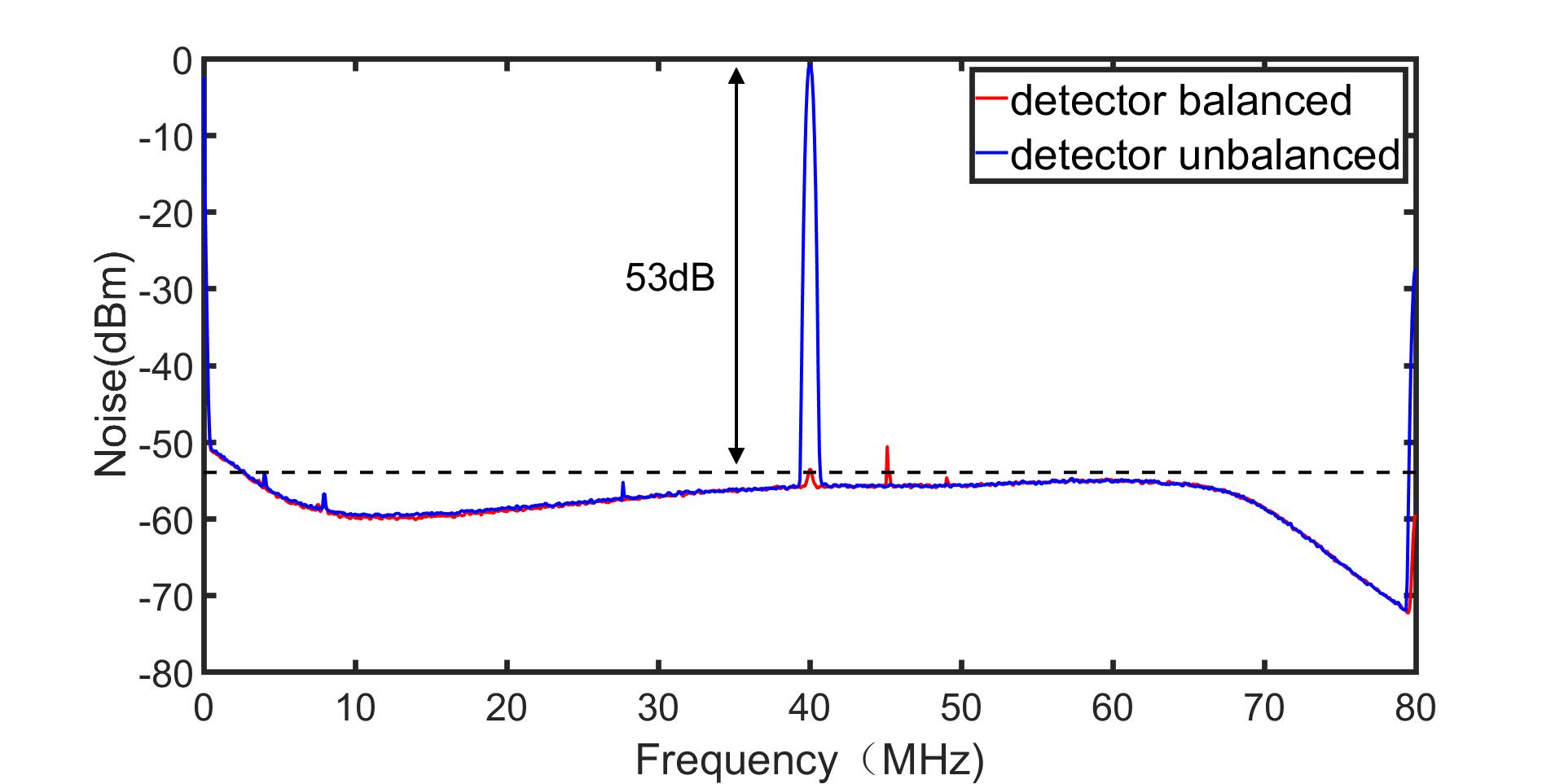}
\caption{Test results of common mode rejection ratio.}
\label{Test results of common mode rejection ratio}
\end{figure}
\par In order to get higher noise clearance between shot noise and electronic noise, a good way is to control the electronic noise and increase the gain[8]. Benefit from our design, we can find an average ratio greater than 10 dB at 1.6 mW and a maximum 16.3 dB can be obtained at 6.8 mW over the 0-70 MHz bandwidth range. This result is better than most high speed detectors such as 13dB at 12mW[9], 7.5dB at 11.3mW[10] or 13dB at 6.4mW[5]. Figure 6 shows the electronic noise and shot noise levels of our detector based on 1550 nm continuous wave laser in different powers.  To test this performance, we modulated the pulsed laser in continuous mode to simulate local oscillator light background and set the signal light to no input as a vacuum state.
\begin{figure}[!t]
\centering
\includegraphics[width=3.2in]{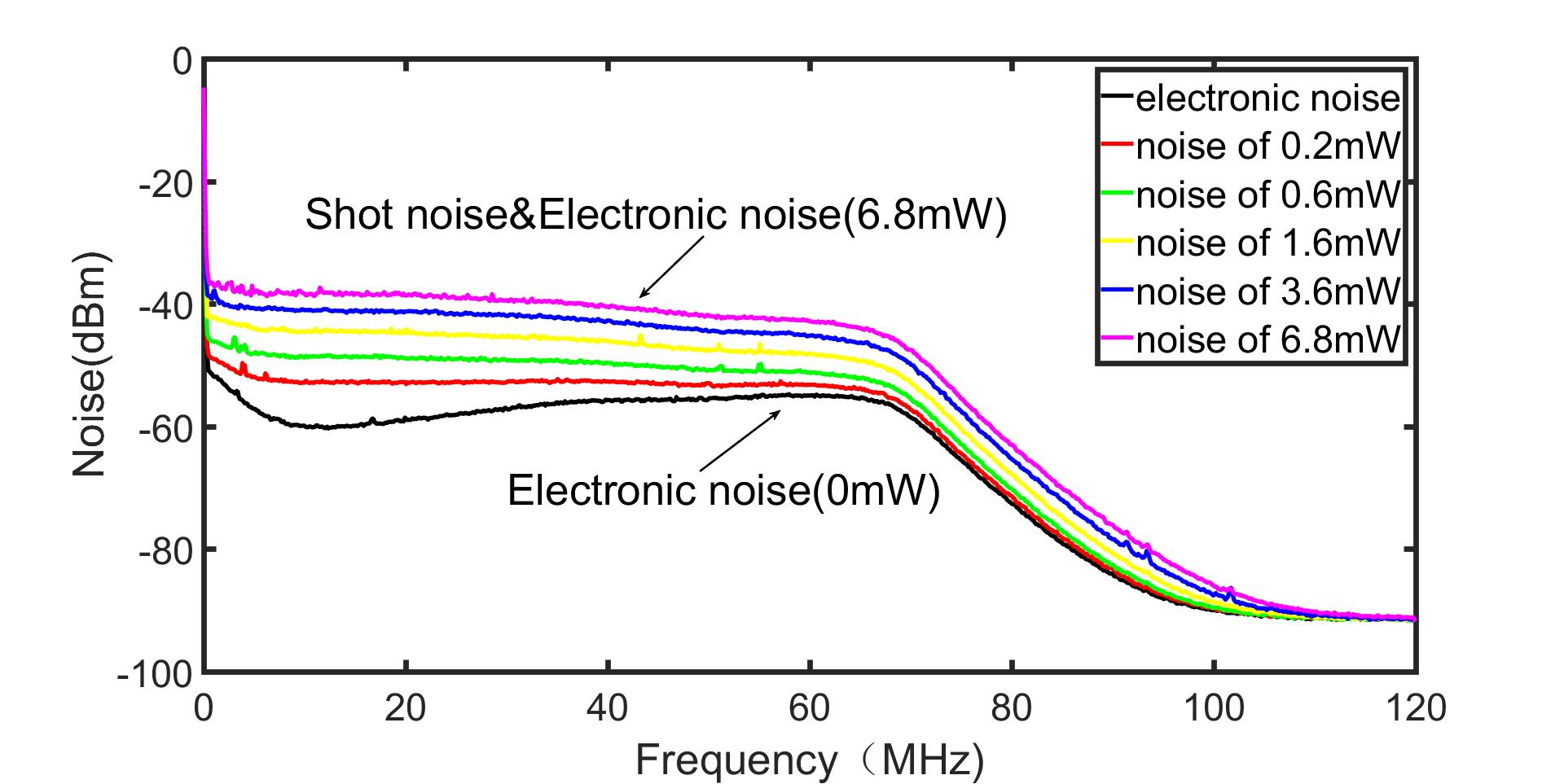}
\caption{Shot noise to electronic noise ratio tests under various LO powers.}
\label{Shot noise to electronic noise ratio under 1.6mW CW laser}
\end{figure}
\par  The methods to increase bandwidth are often in contradiction with those for increasing CMRR and shot noise to electronic noise ratio. Our detector takes into account bandwidth while achieving low noise and high gain, which will be more versatile to CVQKD experiments in complex channel environments. Due to the influence of pulse trailing, the key distribution rate will decrease when the pulse repetition rate of experiment increases and approaches the bandwidth of balanced detector, and the optimal repetition frequency is generally one-third of bandwidth[5], that means our detector can support a 23 MHz repetition rate CVQKD experiment. Besides, in order to prevent the limited detector bandwidth from attenuating signal light, the width of pulse laser is also limited to a minimum of 15 ns[7]. Comparing with the existing
slow CVQKD experiments performed in a stable fiber, our sensitive low-noise balanced detector will be
helpful to achieve a faster CVQKD in complex channel.

\section{Conclusion}
An ultra-sensitive balanced detector based on transimpedance amplifier with low noise for continuous-variable quantum key distribution  is implemented by using low-noise JFET and two-stage amplifier circuit. The NEP of detector is $2.2\ pW/\sqrt{Hz}$ and the gain is 3.2E5 V/W, which lead to a maximum resolution of 16.8 dB in shot noise to electronic noise ratio. Test results show that the detector has a common mode rejection ratio of up to 53 dB and according to CVQKD theory, our detector can support CVQKD experiments based on a maximum pulse repetition rate of 23 MHz and a minimum pulse width of 15 ns.


%





\ifCLASSOPTIONcaptionsoff
  \newpage
\fi

\end{document}